\documentstyle[sprocl,epsfig]{article}

\def\ra{\rightarrow}

\def\beq{\begin{equation}}
\def\eeq{\end{equation}}
\def\bea{\begin{eqnarray}}

\newcommand{\afb}{A_{FB}}
\newcommand{\gv}{\rm GeV}
\newcommand{\pb}{\rm pb^{-1}}
\newcommand{\lsim}{\raisebox{-0.13cm}{~\shortstack{$<$ \\[-0.07cm] $\sim$}}~}
\newcommand{\gsim}{\raisebox{-0.13cm}{~\shortstack{$>$ \\[-0.07cm] $\sim$}}~}

\newcommand{\tb}{\tan\beta}
\newcommand{\non}{\nonumber}
\begin{document}
%%%%%%%%%%%%%%%%%%%%%%%% HEP-PH HEADER %%%%%%%%%%%%%%%%%%%%%%%%%%%
\thispagestyle{empty}
\begin{flushright}
PM/99--51\\
\end{flushright}

\vspace{1cm}

\begin{center}

{\large\sc {\bf SUPERSYMMETRY EFFECTS ON HIGH--PRECISION}}

\vspace*{0.3cm}

{\large\sc {\bf ELECTROWEAK OBSERVABLES}}

\vspace{1cm}

{\large \sc Abdelhak Djouadi} 

\vspace{0.5cm}

Laboratoire de Physique Math\'ematique et Th\'eorique, UMR5825--CNRS,\\
Universit\'e de Montpellier II, F--34095 Montpellier Cedex 5, France.

\vspace*{2cm}

{\large\sc {\bf Abstract}}

\end{center}

\vspace*{0.5cm}

\noindent 

I summarise the virtual effects of the new particles predicted by 
supersymmetric extensions of the Standard Model on the high--precision 
electroweak observables measured at LEP/SLC, the Tevatron and CLEO. I will 
then discuss in some details the two--loop SUSY--QCD corrections to the 
$\rho$ parameter.

\vspace*{3cm}

\centerline{Lecture givent at {\it X Escola Jorge Andr\'e Swieca de 
Part\'{\i}culas e Campos}} 

\centerline{7--12 February 1999, Aguas de Lind\~oia, S\~ao Paulo, Brazil.}

\setcounter{page}{0}

\newpage

%%%%%%%%%%%%%%%%%%%%%%%%%%%%%%%%%%%%%%%%%%%%%%%%%%%%%%%%%%%%%%%%%%
\title{SUPERSYMMETRY EFFECTS ON HIGH--PRECISION \\
ELECTROWEAK OBSERVABLES}

\author{Abdelhak DJOUADI}

\address{Physique Math\'ematique et Th\'eorique, UMR 5825--CNRS, \\
Universit\'e Montpellier II, F--34095 Montpellier Cedex 5, France}

%%%%%%%%%%%%%%%%%%%%%%%%%%%%%%%%%%%%%%%%%%%%%%%%%%%%%%%%%%%%%%
% You may repeat \author \address as often as necessary      %
%%%%%%%%%%%%%%%%%%%%%%%%%%%%%%%%%%%%%%%%%%%%%%%%%%%%%%%%%%%%%%

\maketitle\abstracts{
I summarise the virtual effects of the new particles predicted by 
supersymmetric extensions of the Standard Model on the high--precision 
electroweak observables measured at LEP/SLC, the Tevatron and CLEO. I will 
then discuss in some details the two--loop SUSY--QCD corrections to the 
$\rho$ parameter.}

\vspace*{-5mm}

\section{Introduction/Motivations:} 
Supersymmetry (SUSY) is the most
attractive extension of the Standard Model (SM) \cite{R0,Peter,R1}. It not only
stabilizes the huge hierarchy between the weak and GUT scales against radiative
corrections, but also if SUSY is broken at a sufficiently high scale, it allows
us to understand the origin of the hierarchy in terms of radiative gauge
symmetry breaking. Moreover, SUSY models offer a natural solution to the dark
matter problem and allow for a consistent unification of the all known gauge
couplings. Many new particles are predicted in these theories; the search for
these states and the study of their properties, is and will be, one of the 
major goals of present and future colliders.  

A wide range of searches for supersymmetric particles are performed at present
colliders, in particular at LEP and the Tevatron,  and no direct signal beyond
the SM expectation was observed yet, unfortunately. Therefore new limits on the
masses of these particles, assuming different models, are set by the various
experiments; see for instance Refs.~\cite{R2,pape}. Most of the experimental
limits at LEP2 are close to the kinematical thresholds \cite{flep}, and the 
discovery of SUSY particles has to await for the upgraded Tevatron \cite{ftev},
the LHC \cite{fpp} or for a future $e^+e^-$ linear collider \cite{fee}.  

However, rather than waiting for these future experiments, one could use the
enormous amount of electroweak precision data on  $Z$ and $W$ bosons, collected
at the $e^+ e^-$ colliders LEP and the SLC, and at the Tevatron 
\cite{pape,R3,wolfgang}. 
These high--precision measurements provide a unique tool in the search for
indirect effects of new particles, and in particular SUSY particles, through
possible small deviations of the experimental results from the theoretical
predictions of the minimal SM. This is what I will try to summarize in this 
Lecture.  

In section 2, I will briefly describe the Minimal Supersymmetric
extension of the Standard Model (MSSM), summarise the experimental limits on
the SUSY particle masses and the high--precision measurement of the
electroweak observables. In section 3, I will discuss the observables where
potentially large effects form SUSY particles could be expected: the $Z$ boson
decay widths into hadrons, the $W/Z$ boson self--energies and the radiative
decay $b \to s \gamma$. In section 4, I will focuss on the $\rho$ parameter,
and show an exemple of a calculation of SUSY--QCD corrections at the two--loop 
level, highlightening the new features compared to SM calculations.  
\section{Physical Set--Up}
\subsection{The Minimal Supersymmetric Standard Model}
The MSSM is the most economical low--energy supersymmetric extension of the 
SM. The unconstrained model is defined by the following four basic assumptions
[for more details, see the reviews in Ref.~\cite{R0,Peter}]:

-- It is based on the SM gauge symmetry ${\rm SU(3)_C \times  SU(2)_L \times
U(1)_Y}$. SUSY implies then, that the spin--1 gauge bosons, and their spin--1/2
superpartners the gauginos [bino $\tilde{B}$, winos $\tilde{W}_{1-3}$ and
gluinos $\tilde{g}_{1-8}$] are in vector supermultiplets.  

-- There are only three generations of spin--1/2 quarks and leptons [no
right--handed neutrino] as in the SM. The left-- and right--handed chiral
fields belong to chiral superfields together with their spin--0 SUSY partners
the squarks and sleptons. In addition, two chiral superfields with respective
hypercharges $-1$ and $+1$ for the cancellation of chiral anomalies, are
needed. Their scalar components give separately masses to the isospin +1/2 and
$-$1/2 fermions. Their spin--1/2 superpartners, the higgsinos, will mix with
the winos and the bino, to give the mass eigenstates, the charginos
$\chi_{1,2}^\pm$ and neutralinos $\chi^0_{1,2,3,4}$.  

-- To enforce lepton  and baryon  number conservation, a discrete and 
multiplicative symmetry called R--parity is imposed. The R--parity quantum 
numbers are $R=+1$ for the ordinary particles and $R=-1$ 
for their supersymmetric partners. The conservation of $R$--parity 
has important consequences: the SUSY particles are 
always produced in pairs, in their decay products there is always
an odd number of SUSY particles, and the lightest SUSY particle (LSP)
is absolutely stable.  

These conditions are sufficient to completely determine a globally SUSY
Lagrangian. The kinetic part is obtained by generalizing the notion of 
covariant derivative to the SUSY case, and one then has to add the most 
general superpotential compatible with gauge invariance, renormalizability 
and R--parity. 

To break SUSY, while preventing the reappearance of the quadratic divergences,
one adds to the previous Lagrangian a set of terms which explicitly but softly 
breaks SUSY: mass terms for the gluinos, winos and binos; mass terms for the 
scalar fermions;  mass and bilinear terms for the Higgs bosons; and trilinear 
couplings between sfermions and Higgs bosons.  

This unconstrained MSSM, for generic values of the parameters, might lead to
severe phenomenological problems, such as flavor changing neutral currents
(FCNC), unacceptable amount of additional CP--violation, color and charge
breaking minima, etc...  Furthermore, it contains a huge number of free
parameters, which are mainly coming  from the scalar potential: if we allow for
intergenerational mixing and complex phases, 105 unknown parameters are
introduced in addition to the 19 parameters of the SM! This feature of course
will make any phenomenological analysis a daunting task.  There are,
fortunately, several phenomenological constraints which make some assumptions
reasonably justified to constrain the model. Assuming that there are: no new
source of CP--violation, no FCNC first and second generation sfermion
universality will lead to 19 new input parameters only.  Such a model, with
this relatively moderate number of parameters [especially that, in general,
only a small subset appears when one looks at a given sector of the model] has
much more predictability and is much easier to be discussed phenomenologically.

All the phenomenological problems of the unconstrained MSSM discussed
previously are solved at once if one assumes that the MSSM parameters obey a
set of boundary conditions at the Unification scale.  These assumptions are
natural in scenarii where the SUSY--breaking occurs in a hidden sector which
communicates with the visible sector only through gravitational interactions.
These unification  and universality hypotheses are as follows: unification of
the gaugino masses $m_{1/2}$; universal scalar [sfermion and Higgs boson]
masses $m_0$; universal trilinear couplings $A_0$.  Besides these three
parameters, the SUSY sector is described at the GUT scale by the bilinear 
coupling $B$ and the higgsino mass parameter $\mu$. However, one has to require
that electroweak symmetry breaking  takes place. This results in two 
minimization conditions of the Higgs potential and the equations can be solved 
for $B$ and $|\mu|$. Therefore, in this model, we will have only four
continuous and one discrete free parameters: $\tan \beta$ [the ratio of the
vev's of the two--Higgs doublet fields], $m_{1/2}, m_0, A_0$ and ${\rm sign} 
(\mu)$. This model, usually referred to as the minimal Supergravity model or
mSUGRA, is clearly appealing and suitable for thorough phenomenological 
and experimental scrutinity. 

\vspace*{-2mm}
\subsection{Lower limits on SUSY particle masses} 

The searches for SUSY particles at LEP concern sleptons, stops, sbottoms,
charginos and neutralinos. These various particles decay to SM particles and
two LSPs; therefore, SUSY signatures consist of some combination of jets or/and
leptons and missing energy since the LSP escapes detection. The signal topology
and the background conditions are in practice affected by the SUSY particle and
the LSP mass difference ($\Delta M= m_{\rm SUSY}-m_{\chi^0_1})$ which controls
the visible energy. Since all the background sources are due to well calculable
processes with reasonable production cross sections compared to the signal,
most of the decay channels are studied at LEP2.  Another important key domain
concerns the searches for the MSSM Higgs bosons.  Presently, only the lighter
neutral Higgs bosons $h$ and $A$ can be discovered at LEP2, since the CP--even
Higgs boson $H$ and the charged Higgs bosons $H^\pm$ are expected to be too
heavy.  Each experiment (ALEPH, DELPHI, L3, OPAL) has accumulated data $\sim 55
\; \pb$  at $ \sqrt{s}=183$ GeV in 1997. In october 1998, an amount of $150 \; 
\pb$ at $\sqrt{s}=189$ GeV is already recorded by each experiment, but the 
results including all the statistics for 1998 are not yet available.

At the Tevatron the main sources for SUSY are squarks and gluinos, 
abundantly produced due to the color factors and the strong coupling
constant. Squarks or gluinos are produced in pairs, and decay 
directly or via cascades to at least two LSP's. The classical searches rely 
on large missing transverse energy caused by the escaping LSPs.
In addition, charginos and neutralinos are searched for via their leptonic 
decay channels by  the two Tevatron experiments CDF and D0. Finally, 
searches for the MSSM charged Higgs boson are performed, and
bounds on its mass have been set by both experiments. 
Each experiment has collected an integrated luminosity  of about 110 pb$^{-1}$ 
at $\sqrt{s}=1.8$ TeV. 

Since no evidence for production of supersymmetric particles has been found
at LEP or the Tevatron, experimental limits on their production cross sections 
and masses has been derived. The mass limits are summarised in Table~1 from
Ref.~\cite{R2}. 

\begin{table}
\begin{center}
\begin{tabular}{|c|l|c|l|} 
\hline
Particle & Assumptions & Limit  
 & Exp. source \\ 
\hline \hline
h        & $\tb \geq$0.8  &  78.8              & LEP2 Comb.  \\
A        &$\tb \geq$0.8  &  79.1              & LEP2 Comb.    \\
H$^{\pm}$  &  
Br($\rm H\rightarrow c\bar{s})$ + Br($\rm H\rightarrow c\bar{s}$)=1
& 68  & LEP2 Comb. \\ 
\hline   
\hline
$\tilde{e}_R$ & Br($\tilde{e}_R$ $\rightarrow e \chi_1^0$)=1,
$\Delta$M$\geq 20~\gv$  &  85       & LEP2 Comb.   \\
$\rm \tilde{\mu}_R$  &  BR($\rm \tilde{\mu}_R$ $\rightarrow \mu \chi_1^0$ )=1,
$\Delta$ M $\geq$ 20 $\gv$  &  71       & LEP2 Comb.    \\
$\rm \tilde{\tau}_R$& BR($\rm \tilde{\tau}_R$ $\rightarrow \tau \chi_1^0$)=1,
$\Delta$ M $\geq 20~\gv$  &  75       & LEP2 Comb.   \\
$\tilde{\nu}$    & &  43       & LEP1 $\Gamma_Z$      \\
$\rm \tilde{t}_1$ & BR($\rm \tilde{t}_1$ $\rightarrow  c\chi_1^0$)=1, 
$\Delta$ M $\geq$ 15~$\gv$  &  83       & LEP2 Comb.  \\
 & BR($\rm \tilde{t}_1$ $\rightarrow  c\chi_1^0$ )=1,
$ {\rm LEP2;~m_{\chi_1^0}  \leq  50~GeV}$  &  122       & CDF     \\
$\rm \tilde{b}_1$    & BR($\rm \tilde{b}_1$ $\rightarrow b\chi_1^0$ )=1, 
$\Delta$ M $\geq  15~\gv$  &  75       & LEP2 Comb.  \\
%          &                           &   &                  \\   
$\rm \tilde{q}$    & $\rm m_{\tilde{q}} \geq m_{\tilde{g}}$
 &  216       & CDF     \\ 
&  $\rm m_{\tilde{q}} \geq m_{\tilde{g}}$  &  260       & D0     \\
\hline
\hline
%          &                           &   &                  \\   
$\rm \tilde{g}$    & $\mu \leq -100$ or $\mu \geq 200~\gv$ 
 &  173       & CDF     \\ 
 & $\mu \leq -100$ or $\mu \geq 200~\gv$ &  187       & D0     \\
$\chi^{\pm}_{1}$ & Higgsino $\Delta$M $\geq 3~\gv$ & 63 & LEP2 Comb.\\ 
 & Gaugino $\rm m_0$ $\geq 200~ \gv$ & 94.3 & LEP2$^*$ \\ 
$\chi^0_{1}$ & Large $\rm m_0$ &  32.5  &   LEP2$^*$           \\ 
\hline 
\end{tabular}
\end{center}
\caption{Summary of lower mass limits obtained at the Tevatron and   
at LEP2 up to $\sqrt{s}$ = 183, 189 GeV [$^*$ at which 
just a small fraction of data has been used]; From Ref.~[4].}
\vspace*{-0.5cm}
\end{table}

\subsection{High--precision data} 

The $e^+ e^-$ colliders LEP and the SLC, in operation since 1989, have
collected an enormous amount of electroweak precision data on  $Z$ and $W$
bosons.  Measurements at the $Z$--pole of $Z$ boson partial and total decay
widths, polarisation and forward--backward asymmetries where made at the
amazingly high accuracy of the level of one percent to one per mille; see Table
2 from Ref.~\cite{wolfgang}.  The $W$ boson properties have in parallel been
determined at the $p \bar{p}$ collider Tevatron with a constant increase in
accuracy. The ongoing experiments at LEP2 and the near-future Tevatron upgrade
will also, in the coming years, provide us with further increase in precision,
in particular on the mass of the $W$ and the SLC might continue to improve the 
impressive accuracy already obtained in the electroweak mixing angle.

Another measurement of interest here, is the branching ratio for the radiative 
flavor changing decay $B \to X_s \gamma$ performed at LEP1 and by the CLEO
collaboration. The most precise value and the SM expectation for the
branching ratio are \cite{CLEO}: 
\begin{eqnarray}
{\rm BR} (B \to X_s \gamma) \times 10^4 = [3.14 \pm 0.48]_{\rm exp} \ \ 
{\rm and} \ \ [3.29 \pm 0.33]_{\rm th} \nonumber
\end{eqnarray}

The availability of both highly accurate measurements and theoretical
predictions, at the level of 0.1\% precision and better, provides tests of the
quantum structure of the SM, thereby probing its still untested scalar sector,
and simultaneously accesses alternative scenarios such as the supersymmetric
extension of the SM. Indeed,  the lack of direct signals from new physics
beyond then makes the high-precision experiments a unique tool also in the
search for indirect effects, through possible small deviations of the
experimental results from the theoretical predictions of the minimal SM.

\begin{table}[t]
\caption[]   {Precision observables: 
              experimental results from combined
              LEP and SLD data for $Z$ observables and 
              combined $p\bar{p}$
              and LEP data for $M_W$, with the SM predictions;
              From Ref.~[11].}
\vspace{0.5cm}
\begin{center}
\begin{tabular}{| l  l  r | }
\hline 
Observable & Exper. value  & SM  best fit \\
\hline 
  & &  \\[-0.3cm]
$M_Z$ (GeV) & $91.1867\pm0.0019$ & 91.1865    \\
%\hline
$\Gamma_Z$ (GeV) & $2.4939\pm 0.0024$ & 2.4956 \\
%\hline
$\sigma_0^{had}$ (nb) & $41.491\pm 0.058$ & 41.476 \\
%\hline
 $ R_{had}$ & $20.765\pm 0.026 $ & 20.745 \\
%\hline
$R_b$  & $0.21656\pm 0.00074$ & 0.2159 \\
%\hline
$ R_c$  & $0.1732\pm0.0048$ & 0.1722 \\
%  \hline
$\afb^{\ell}$ & $0.01683 \pm 0.00096$ & 0.0162 \\
%\hline 
$\afb^b$ & $0.0990 \pm 0.0021$ &  0.1029 \\
%\hline
$\afb^c$ & $0.0709 \pm 0.0044$ &  0.0735 \\
%\hline 
$A_b$            & $0.867\pm 0.035$  & 0.9347 \\
%\hline
$A_c$            & $0.647\pm 0.040$  & 0.6678 \\
%\hline
$\Delta \rho$ & $1.0041\pm 0.0012$ &  1.0051 \\
%\hline
$\sin^2\theta_W$  & $0.23157\pm 0.00018$ & 0.23155\\
%\hline
%$\sin^2\theta_W (A_{LR})$ & $0.23049\pm 0.00050$ & $0.2317\pm 0.0012$   \\
% LEP$+$SLC   &  $0.23143\pm 0.00028$    &                    \\
%\hline
$M_W$ (GeV) & $80.39 \pm 0.06$ & 80.372 \\[0.1cm]
%  & &  \\
\hline
\end{tabular}
 \vspace{0.5cm}
\end{center}
%  \vspace{-1.5cm}
%\clearpage
\end{table}

\section{Potentially large virtual SUSY effects} 

\subsection{The $\rho$ parameter}

A possible SUSY signal might come from the contribution of the SUSY 
particle loops to the electroweak gauge--boson self--energies \cite{R6,manuel}:
if there is a large splitting between the masses of these particles,
the contribution will grow with the mass of the heaviest particle and 
can be sizable. This is similar to the SM case, where the top/bottom 
weak isodoublet generates a quantum correction that grows as $m_t^2$. 
This contribution enters the electroweak observables via the $\rho$ 
parameter~\cite{R7}, which measures the relative strength of the neutral 
to charged current processes at zero momentum--transfer. It is mainly from 
this contribution that the top--quark mass has been successfully predicted 
from the measurement of $\sin^2\theta_W$ and $M_W$, a triumph for the 
electroweak theory. 

The $\rho$ parameter, in terms of the transverse parts of 
the $W$-- and $Z$--boson self--energies at zero momentum--transfer, 
is given by 
\beq
\rho = (1-\Delta \rho)^{-1}\ ; \ \Delta \rho = 
\Pi_{ZZ}(0)/ M_Z^2 - \Pi_{WW}(0)/M_W^2  
\eeq
In the SM, the contribution of a fermion isodoublet $(u,d)$ to $\Delta \rho$ 
reads at one--loop: 
\beq
\Delta \rho_0^{\rm SM} = \frac{N_c G_F}{8 \sqrt{2} \pi^2} F_0 
\left( m_u^2, m_d^2 \right) \ , 
F_0(x,y)= x+y - \frac{2xy} {x-y} \log \frac{x}{y} 
\eeq
The function $F_0$ vanishes if the $u$-- and $d$--type quarks are 
degenerate in mass: $F_0(m_q^2, m_q^2)=0$; in the limit of large quark 
mass splitting it becomes proportional to the heavy quark mass 
squared: $F_0(m_q^2,0)=m_q^2$. Therefore, in the SM the only relevant
contribution is due to the top/bottom weak isodoublet. Because 
$m_t \gg m_b$, one obtains $\Delta \rho ^{\rm SM}_0 = 3 G_Fm_t^2/(8 
\sqrt{2} \pi^2)$, a large contribution which allowed for the prediction
of $m_t$. However, in order that the predicted value agrees with the 
experimental one, QCD corrections have to be included. These two--loop 
corrections have been calculated ten years ago, leading to a 
result~\cite{R7a}:
$\Delta \rho ^{\rm SM}_1 = - \Delta \rho_0^{\rm SM} \cdot 
\frac{2}{3} \frac{\alpha_s}{\pi} (1+\pi^2/3 )$. For the value $\alpha_s 
\simeq 0.12$, the QCD correction \cite{R7b} decreases the one--loop result 
by approximately $10\%$ and shifts $m_t$ upwards by an amount of $\sim 10$~GeV. 

In the SUSY extension of the SM, additional contributions might come from the 
additional Higgs bosons, the charginos and neutralinos as well as well from
the scalar fermions which could possibly have large mass splittings. 
 
The first set of possible contributions might be due to the extended Higgs
sector which leads to a quintet of scalar [two CP--even $h$ and $H$, a
pseudoscalar $A$ and two charged $H^\pm$] particles \cite{R1}. While a strong
upper bound on the mass of the light Higgs boson $h$ can be derived, $M_h \lsim
130$ GeV, the heavy neutral $H$, $A$ and charged $H^\pm$ Higgs bosons may have
masses of the order of the electroweak symmetry scale up to about 1~TeV
\cite{Peter}.  In a general two--Higgs doublet model, the masses of these Higgs
bosons are not related, and large mass splitting between the particles might be
present, leading to possible large contributions to the $\rho$ parameter
\cite{R8}.  In the MSSM, however, the Higgs system is strongly constrained and
is described by two parameters [up to radiative corrections which will not
alter the discussion]: one mass parameter which is generally identified with
the pseudoscalar $A$ mass, $M_A$ and $\tan \beta$. The Higgs boson masses and
couplings to gauge bosons are related in such a way that they lead to large
cancellations in their contributions to the $\rho$ parameter \cite{manuel}. For
instance, in the decoupling limit where the $A$ boson mass is large compared to
$M_Z$, the heavy Higgs bosons are nearly mass degenerate, $M_H \simeq M_{H^\pm}
\simeq M_A$, and their couplings to gauge bosons tend to zero, while the
lightest CP--even $h$ particle reaches its maximal mass value, and has almost
the same properties as the SM Higgs particle. The contribution of the Higgs
sector of the MSSM to the $\rho$ parameter is then practically  the same as in
the SM, i.e. giving rise to logarithmic $\log M_h/M_Z$ effects which are rather
small \cite{pape}.  

The two charginos and the four neutralinos of the MSSM might also have large
mass splittings. In fact, when the higgsino mass parameter $\mu$ is large
compared to the gaugino masses, one has the mass hierarchy: $2 m_{\chi_1^0} \sim
m_{\chi_2^0} \sim m_{\chi_1^+} \gg m_{\chi_3^0} \sim m_{\chi_4^0} \sim
m_{\chi_2^+}$, and one would expect large contributions to $\Delta \rho$.
However, a close inspection of the mass matrices for neutralinos and charginos,
shows that the only terms which could break the custodial SU(2) symmetry, and
hence contribute to the $\rho$ parameter, are proportional to $M_W$ only. This
leads to a contribution \cite{manuel} $\Delta \rho \lsim 10^{-4}$ which is much
too small to be detected; see Table 2.

\begin{figure}[htb]
\begin{center}
\mbox{
\psfig{figure=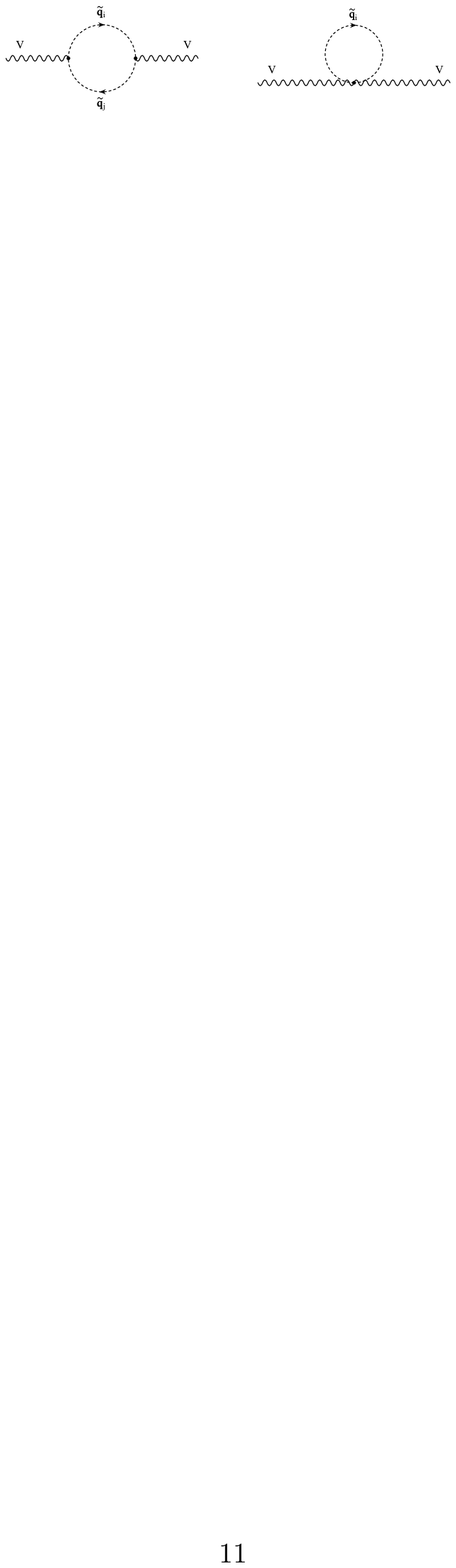,width=8cm,bbllx=210pt,bblly=680pt,bburx=397pt,bbury=720pt}}
\end{center}
\caption[]{Diagrams for the contribution of squarks
to the gauge boson self--energies at one--loop.}
\end{figure}

Another possible contribution might come from the scalar partners of each SM 
fermions, $\tilde{f}$. The current eigenstates, $\tilde{f}_L$ and $\tilde{f}_R$,
mix to give the mass eigenstates. The mixing angle is proportional to the 
fermion mass and therefore is important only in the case of the third 
generation scalar fermions. 
In particular, due to the large value of $m_t$, the mixing angle 
$\theta_{\tilde{t}}$ between $\tilde{t}_L$ and $\tilde{t}_R$ can be 
very large and lead to a scalar top quark  $\tilde{t}_1$ possibly much 
lighter than the $t$--quark and all the scalar partners of the light 
quarks. The mixing in the $\tilde{b}$--squark sector can be
sizable only in a small area of the SUSY parameter space. Neglecting this 
mixing, $\Delta \rho$ is given at one--loop order by the simple expression
$[s_t=\sin \theta_{\tilde{t}}, c_t = \cos \theta_{\tilde{t}}$]
\begin{eqnarray}
\Delta \rho ^{\rm SUSY}_0 = \frac{3 G_F}{8 \sqrt{2} \pi^2} \left[ -
s_t^2 c_t^2 F_0\left( m_{\tilde{t}_1}^2,  m_{\tilde{t}_2}^2 \right)
+ c_t^2 F_0\left( m_{\tilde{t}_1}^2, m_{\tilde{b}_L}^2 \right) + s_t^2 
F_0\left(m_{\tilde{t}_2}^2,  m_{\tilde{b}_L}^2 \right) \right] 
\end{eqnarray}
As can be seen from $F_0$ in eq.~(2), the contribution of a scalar 
quark doublet vanishes if all masses are  degenerate. This means that 
in most SUSY scenarios, where the scalar partners of the leptons and 
light quarks are in general 
almost mass degenerate, only the third generation will contribute.
In a large area of the parameter space, the scalar top mixing angle is 
either very small $\theta_{\tilde{t}} \sim 0$ or maximal, $\theta_{\tilde{t}} 
\sim -\pi/4$. 
The contribution $\Delta \rho_0^{\rm SUSY}$ is shown in Fig.~2
as a function of the common scalar mass $m_{\tilde{q}}=m_{\tilde{t}_{L,R}}
=m_{\tilde{b}_{L}}$ for $\tan \beta=1.6$ in these two scenarios 
[$m_{\rm LR}=0$ and 200 GeV, respectively, where $m_{\rm LR}$ is the 
off--diagonal term in the $\tilde{t}$ mass matrix]. The contribution can be at 
the level of a few per mile and therefore within the range of the experimental
observability. Relaxing the assumption of a common scalar quark mass, 
the corrections can become even larger~\cite{R6,manuel}. 

\begin{figure}[htb]
\begin{center}
\mbox{
\psfig{figure=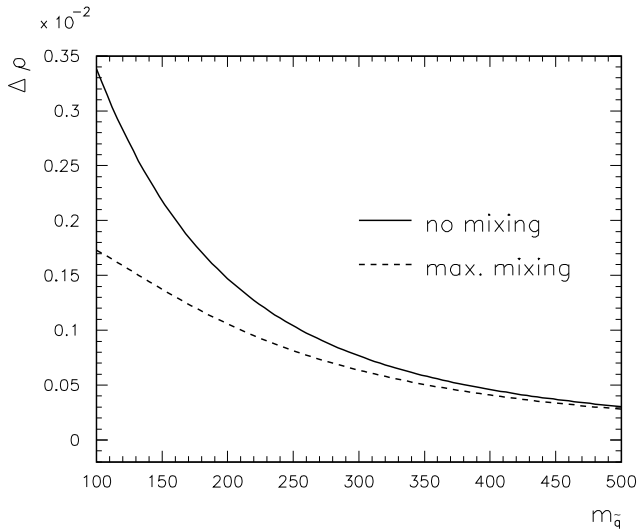,width=8cm,height=6.5cm,bbllx=140pt,bblly=285pt,bburx=450pt,bbury=535pt}}
\end{center}
\caption[]{One--loop contribution of the $(\tilde{t}, \tilde{b})$ 
doublet to $\Delta \rho$ as a function of the common mass $m_{\tilde{q}}$, 
for $\theta_{\tilde{t}} =0$ and $\theta_{\tilde{t}} \sim-\pi/4$ 
[with $\tan \beta=1.6$ and $m_{\rm LR}=0$ and 200 GeV, respectively.} 
\end{figure}

\subsection{$Z$ boson decays into hadrons} 

Because of the large value of $\alpha_s$, the potentially largest SUSY
effects are expected to come from corrections involving strong interactions.  
One of the simplest cases where SUSY--QCD corrections can be looked for is 
the cross section for $e^+ e^- \rightarrow {\rm hadrons}$. In addition to the 
standard corrections, virtual gluon exchange and gluon emission in the final 
state, one has also diagrams where squarks and gluinos are exchanged in the 
loops \cite{eehad}. Unfortunately, because gluinos and squarks are expected to 
have masses above $\sim 200$ GeV, the corrections are rather small at present 
energies. For instance, for the hadronic width of the $Z$ boson, $\Gamma(Z 
\rightarrow \bar{q} q)$ , the SUSY--QCD correction is less than 0.2\% for 
realistic values of $m_{\tilde{g}}$ and $m_{\tilde{q}}$, which is less than the 
experimental accuracy of the measurement.

Another source of possible large SUSY effects is the $Z b\bar{b}$ vertex which
can alter the SM prediction for the decay width $\Gamma(Z\ra b \bar{b})$ or
$R_b$ which is measured with an accuracy of a few per mile; see Table 2. 
Indeed, the couplings of the Higgs bosons to bottom and top quarks are,
respectively, proportional to $m_b \tan \beta$ and $m_t/\tan \beta$ so that the
couplings to $b(t)$ quarks are enhanced for large (small) values of $\tan
\beta$. The exchange \cite{zbbhiggs}  of the light $h,A$ bosons with $b$-quark
loops for high--$\tan \beta$ and/or the charged $H^+$ boson with $t$--quark
loops for low $\tan \beta$ might lead to large contributions to the $Z
b\bar{b}$ vertex.  In addition, chargino and top squarks can be exchanged in 
the $Z b\bar{b}$ vertex \cite{zbbstop}, and since the $\chi_1^\pm \tilde{t} b$
coupling is also proportional to $m_t/\tan \beta$, large contributions can
occur in the vertex.

These two corrections were extensively discussed in the context of the $R_b$
crisis \cite{Rbcrisis} a few years ago, when this quantity showed a 4$\sigma$ 
deviation from the SM expectation. However, with the present experimental bounds
on the Higgs bosons, charginos and top  squark masses, these corrections 
are now too small to be detectable with the present accuracy on $R_b$. 

\subsection{The decay $b \to s\gamma$} 

In the SM, the radiative and flavor changing decay $b \to s\gamma$ is mediated
by loops where top quarks and $W$--bosons are exchanged. In SUSY theories
\cite{bsgamma}, additional contributions are provided by loops of charginos and
stops and loops of top quarks and charged Higgs bosons. Since both SM and SUSY
contributions appear at the same level of perturbation theory, the measurement
of the inclusive decay $B \ra X_s \gamma$, turns out to be a very powerful tool
to constrain the MSSM.  

Indeed, assuming for instance that the stop/chargino loops as absent, as
is the case in a general two--Higgs doublet model, the charged Higgs boson
mass can be strongly constrained from $H^\pm$/top contribution to the decay.
The present experimental value given by CLEO \cite{CLEO}, implies for instance 
$M_{H^\pm} \gsim 260$ GeV. In the MSSM, a possible negative interference
between the chargino/stop and the $H^\pm$ boson loops might take place,
leaving out only the SM contribution. This happens only in some areas
of the MSSM parameter space; in other areas, SUSY loop contributions can 
generate unbearable effects in the decay. For instance, for an $H^\pm$
mass of the order of 100 GeV, the sum of the chargino and the stop mass 
should be smaller than $\sim 400$ GeV in order not to violate the experimental
bound on the decay; the measurement can thus lead to interesting and strong 
constraints.  

\subsection{Global Fits} 

Since for the time being, no deviation from SM expectations has been 
identified, the presence of SUSY particle contributions to the precision 
observables can be exploited to constrain the allowed range of the MSSM 
parameters. This can be done by looking at specific observables in which only
a small set of SUSY parameters enters, but one can also make a global fit of 
the complete set of data  where all SUSY parameters are involved.
Due to the proliferation of the unknown parameters, this is rather difficult to 
achieve in the unconstrained MSSM. In the mSUGRA model, however, thanks to the
restricted set of input parameters, one can obtain interesting constraints;
for recent reviews, see Refs.~\cite{fits,kaoru}. 

In Ref.~\cite{kaoru} that we will closely follow here, a global fit in the
mSUGRA scenario has been performed. The output is shown in Fig.~3 in the
$m_0$--$m_{1/2}$ plane for three values of $\tan \beta=2,10,35$ and both 
signs of
$\mu$; the remaining mSUGRA parameter $A_0$ was allowed to vary between $-500$
GeV $\leq A_0 \leq +500$ GeV. The main ingredients of the analysis are as
follows \cite{kaoru}:  

$i)$ Require correct EW symmetry breaking, that the SUSY particle masses are 
physical, and that the lightest neutralino is indeed the LSP; this leads to
a disallowed region in Fig.~3 in the upper left corner [solid lines]. 

$ii)$
Impose  the experimental lower bound on the lightest chargino 
mass $m_{\tilde{\chi}^\pm_1} > 91$~GeV; this then 
excludes the region below the horizontal solid line. 

$iii)$ Include the SUSY corrections to the
$W/Z$ boson self--energies and in particular to $\Delta \rho$ [the vertex 
corrections are very small and do not lead to any constraint] and perform a 
fit;  the region from the solid line to the dashed contour is then excluded at 
the 95\% confidence level.  

$iv)$ Include the SUSY contribution to ${\rm BR}(B \to 
X_s\gamma)$; the region from the dashed to the dotted contours is 
the excluded at the 95\% confidence level.  

The portion of the $m_0$--$m_{{1/2}}$ plane which is above and to the right of 
all the contours is the favored region for the mSUGRA scenario. One can
see that the constraint from the chargino mass bound is significant. The 
constraint from the $\rho$ parameter excludes a corner of the $m_0$--$m_{{1/2}}$
plane corresponding to small values of $m_0$ and $m_{{1/2}}$. The constraint
from ${\rm BR}(B\to X_s\gamma)$ is the most significant one and excludes
a large portion of the $m_0$--$m_{{1/2}}$ plane for ${\rm sign}(\mu) < 0$ or 
$\tan\beta$ large; the constraint become very strong when both conditions are 
met.

\smallskip

In order to treat the SUSY loop contributions to the electroweak
observables at the same level of accuracy as the standard contributions,
higher--order corrections should be incorporated; in particular the QCD 
corrections, which because of the large value of the strong coupling 
constant can be rather important, must be known. 

Recently the next--to--leading order QCD correction to the decay $b \to 
s\gamma$ have been completed \cite{bsgammaQCD} in both a general two--Higgs 
doublet model [i.e. the correction to the contribution of the $H^\pm$ boson] 
and in the SUSY case [i.e. including the stop/chargino loops]. The correction
turns out to be rather important, and must be taken into account. 
Also recently, the results for the ${\cal O}(\alpha_s)$ correction to the 
contribution of the scalar top and bottom quark loops to the gauge boson 
self--energies and hence to the $\rho$ parameter have been derived
\cite{drho,drall}. In the next section, I will summarize the main features
and main results of this calculation. 

%%%
\begin{figure}[htbp]
\begin{center}
\leavevmode\psfig{figure=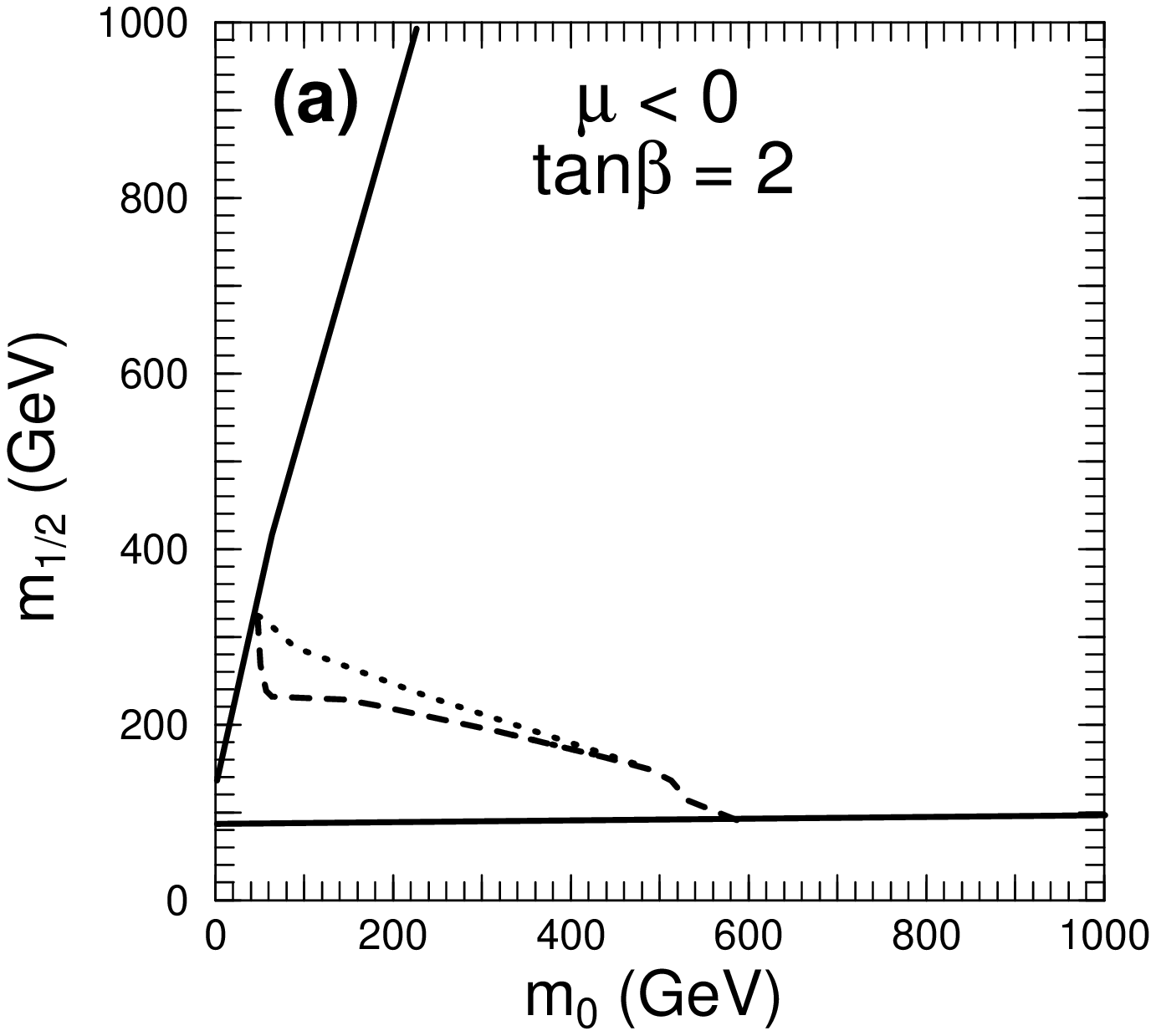,width=5.5cm,silent=0}
\leavevmode\hspace*{0.5cm}
\leavevmode\psfig{figure=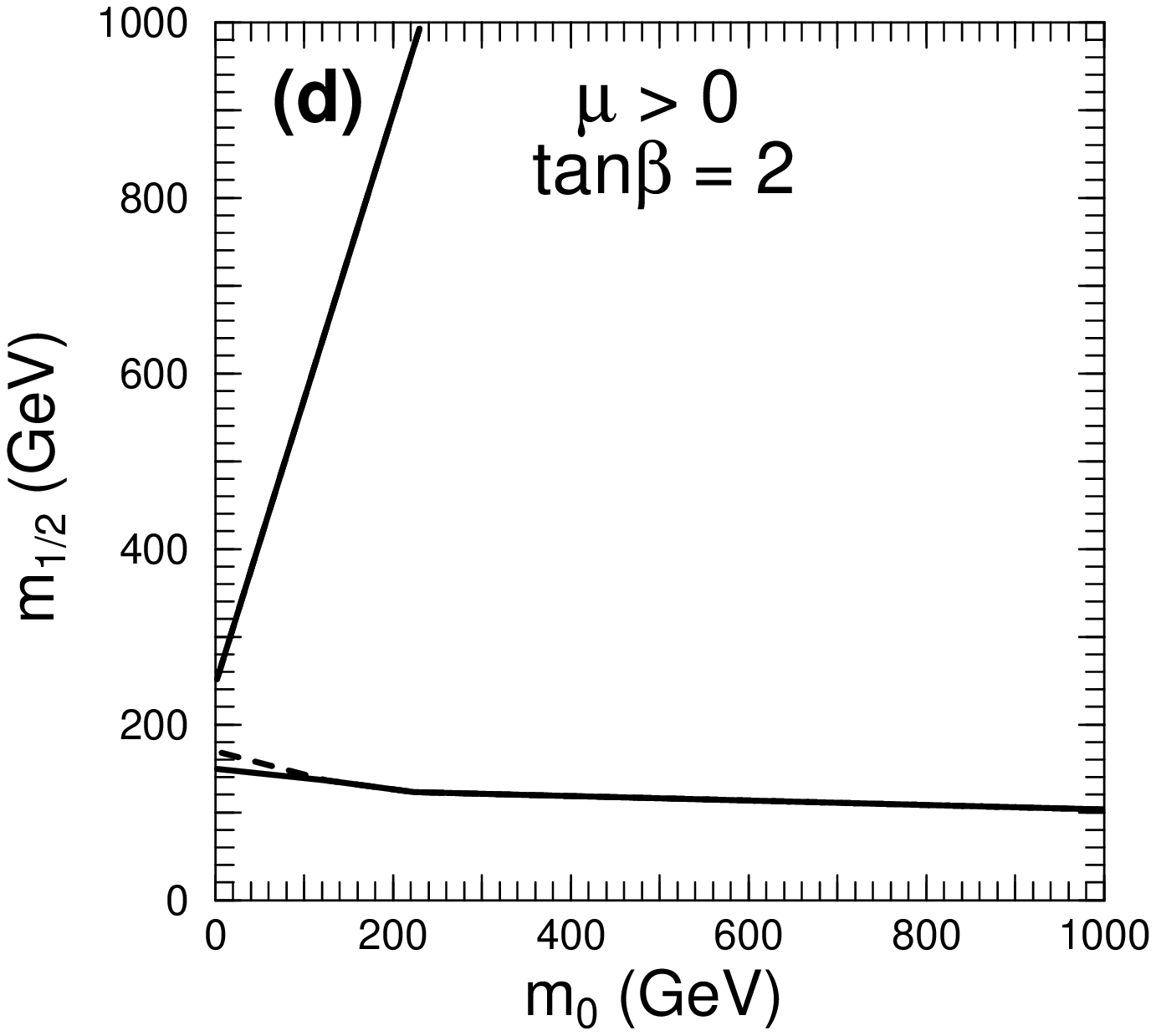,width=5.5cm,silent=0}
\end{center}
\vspace*{0.5cm}
\begin{center}
\leavevmode\psfig{figure=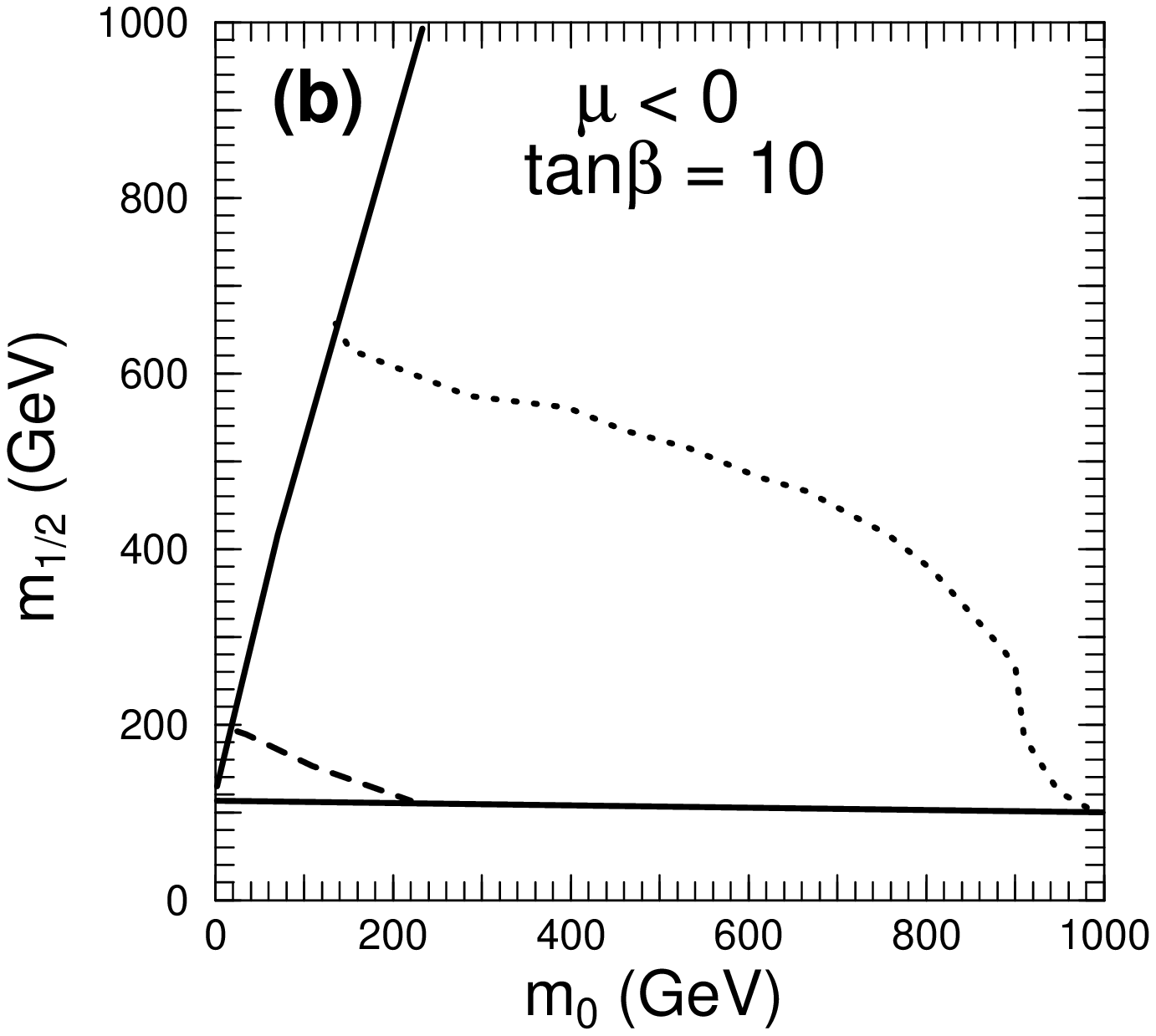,width=5.5cm,silent=0}
\leavevmode\hspace*{0.5cm}
\leavevmode\psfig{figure=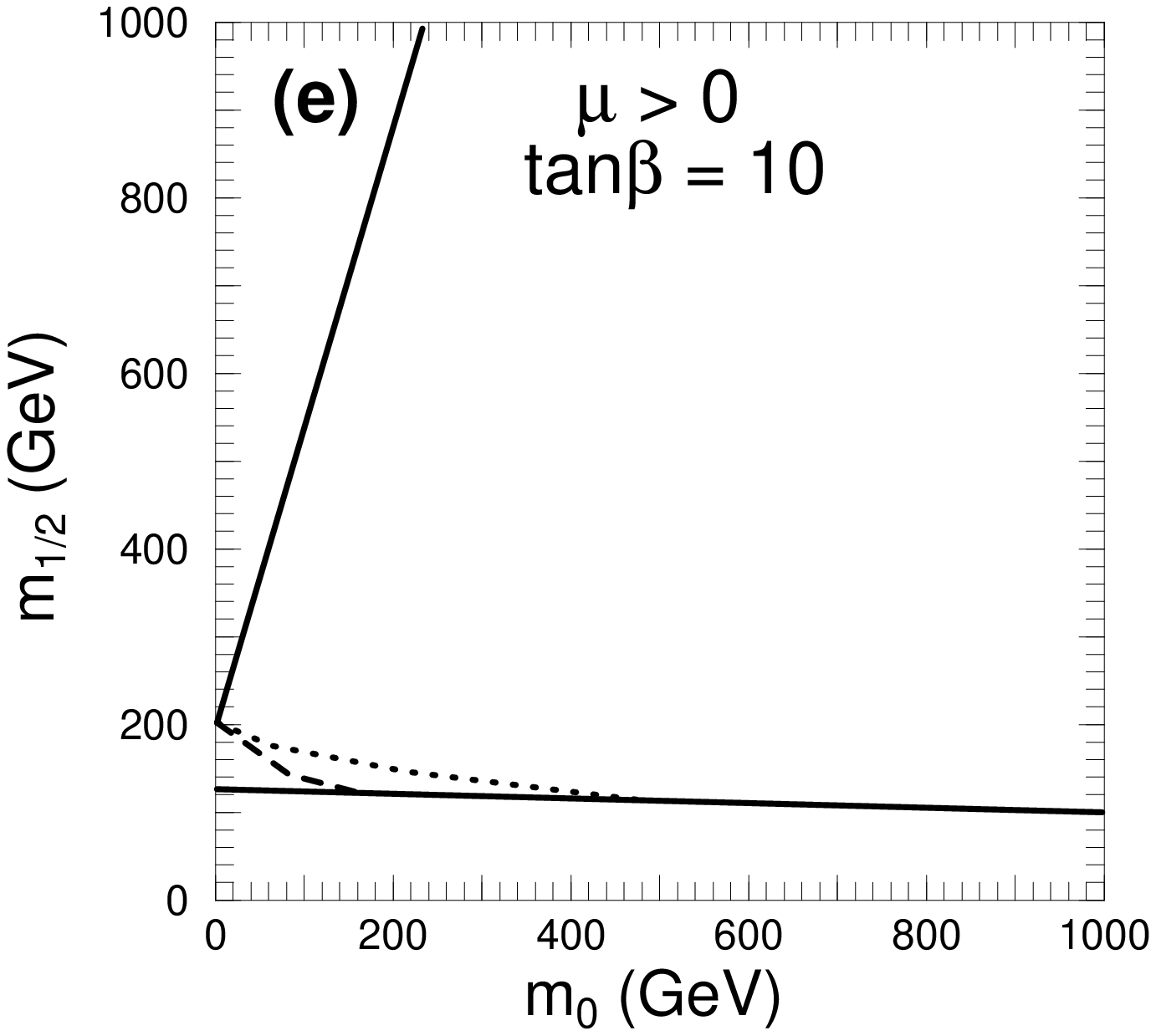,width=5.5cm,silent=0}
\end{center}
\vspace*{0.5cm}
\begin{center}
\leavevmode\psfig{figure=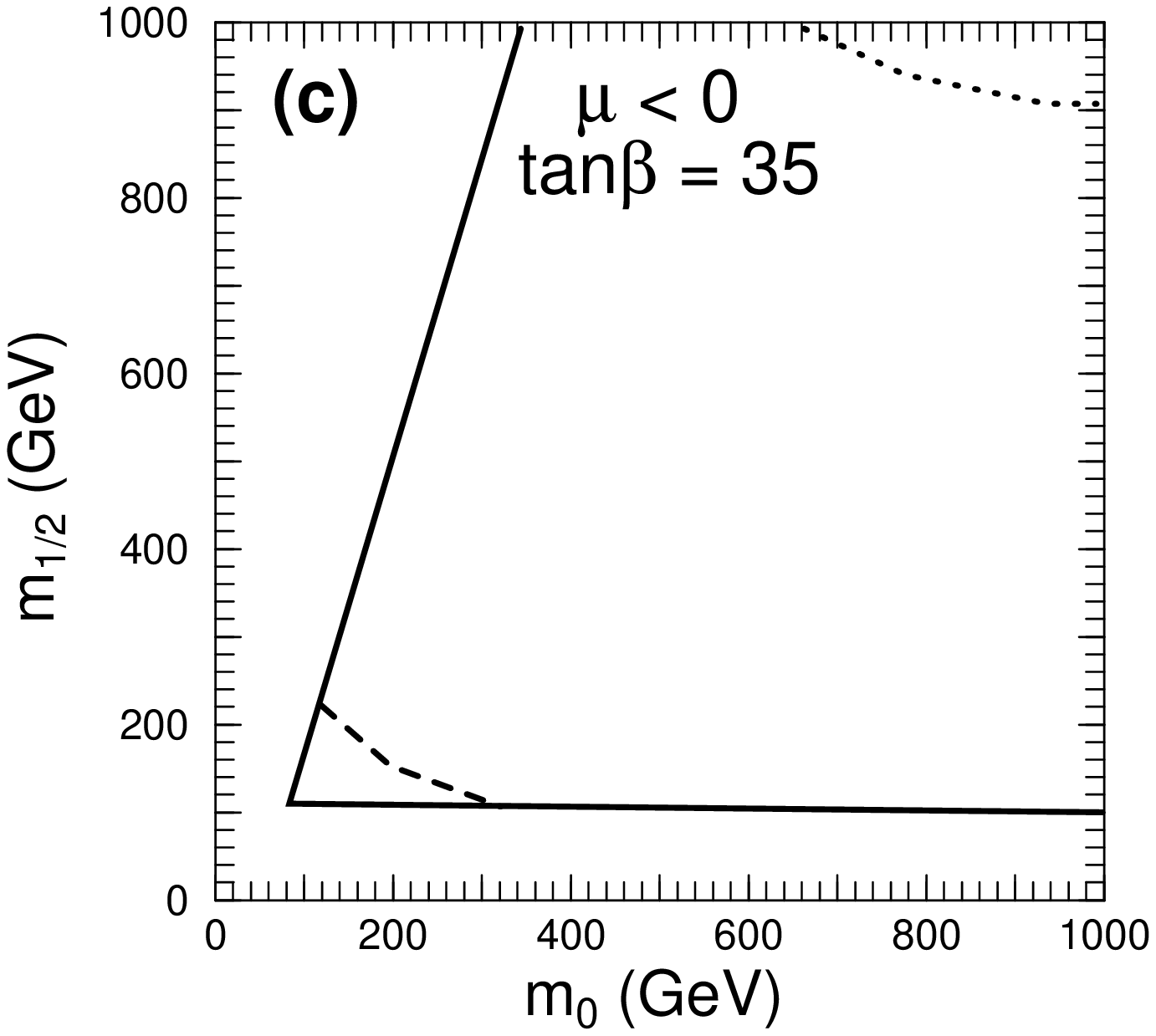,width=5.5cm,silent=0}
\leavevmode\hspace*{0.5cm}
\leavevmode\psfig{figure=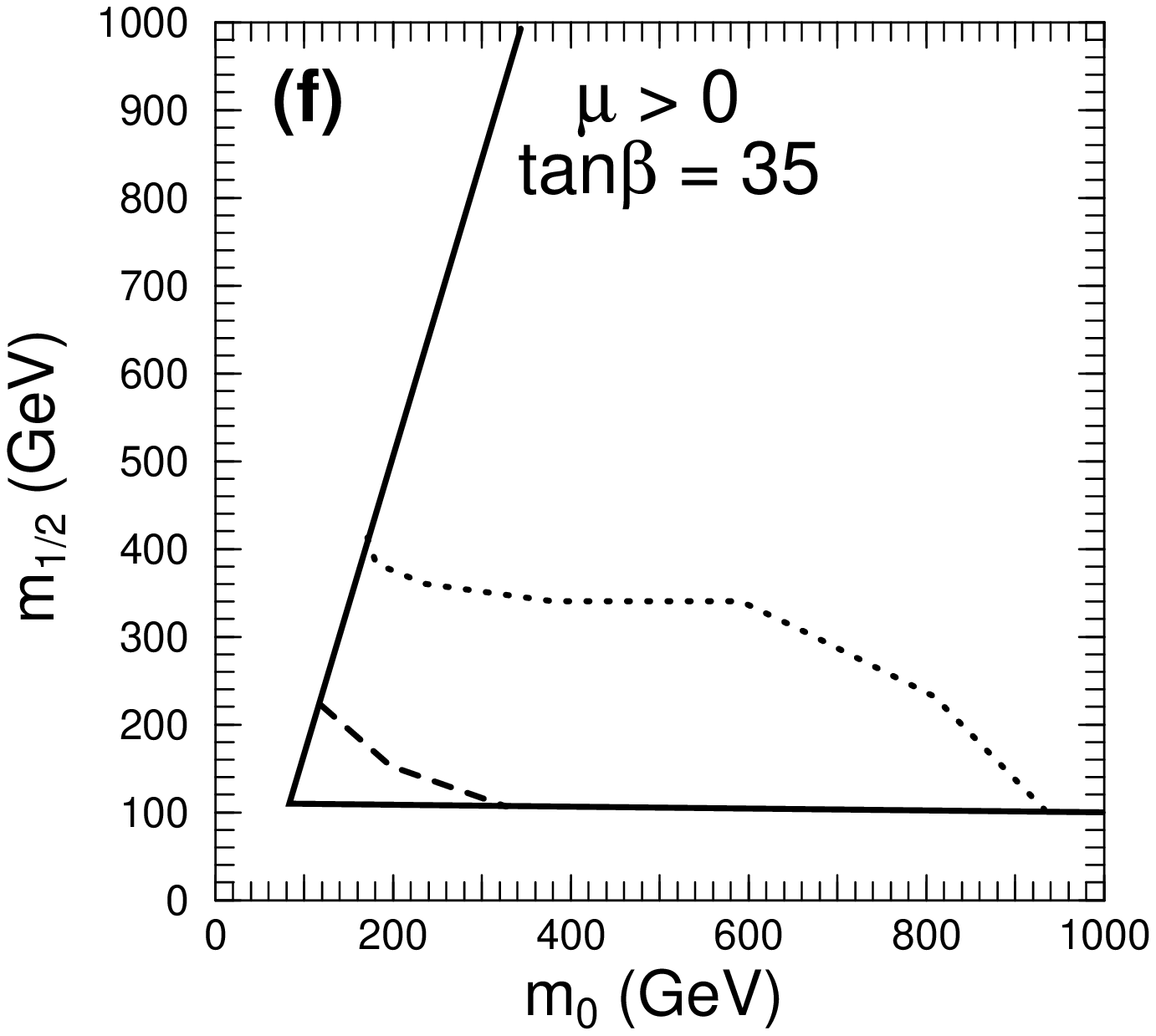,width=5.5cm,silent=0}
\end{center}
\caption{Favored regions in the mSUGRA $m_0$--$m_{1/2}$ plane lie in the 
region which is above and to the right of all drawn contours; From Ref.~[25].}
\end{figure}

\section{QCD corrections to the $\rho$ parameter} 

\subsection{The SUSY--QCD Lagrangian}

In SUSY theories, as strongly interacting 
particles one has in addition to gluons and quarks, the gluinos $\tilde{g}$ 
and three generations of left-- and right--handed squarks, $\tilde{q}_L$ and 
$\tilde{q}_R$. The interactions between gluon [$V^\mu$], gluino 
$[\lambda]$, quark [$\psi_i$] and scalar quark $[\phi_i$] fields are dictated 
by SU(3)$_{\rm C}$ gauge invariance and are given by the Lagrangian
\begin{equation}
{\cal L} = {\cal L}_{\rm kin} + {\cal L}_{\rm mat} + {\cal L}_{\rm self}+
{\cal L}_{\rm Yuk}+ {\cal L}_{\rm soft} 
\end{equation}
There is first the self--interactions of the gauge fields, where in addition
to the 3 and 4--gluon vertices that we do not write, there is a 
term containing the interaction of the gluinos with the gluons [$\sigma^\mu$ 
are the Pauli matrices which help to write down things in a two-component 
notation and $f_{abc}$ the structure constants of SU(3)]: 
\begin{eqnarray}
{\cal L}_{\rm kin}= ig f_{abc} \lambda^a \sigma^\mu \bar{\lambda}^b V_\mu^c
+``3V" + ``4V" 
\end{eqnarray}
Then, there is a piece describing the interaction of the gauge and matter
particles
\begin{eqnarray}
{\cal L}_{\rm mat} &=& -g T^a_{ij}V_\mu^q \bar{\psi}_i \bar{\sigma}^\mu
\psi_j  -ig T^a_{ij} V_\mu^q \phi_i^* \partial^{\leftrightarrow} \phi_j
+ g^2 (T^a T^b)_{ij} V^a_\mu V^{\mu b} \phi_i^* \phi_j  \non \\
&& +ig_Y \sqrt{2} T^a_{ij} (\lambda^a \psi_j \phi_i^*- \bar{\lambda}^a 
\bar{\psi}_i \phi_j) 
\end{eqnarray}
Besides the usual term  for the gluon--quark interaction and the terms 
for purely scalar QCD [the derivative term for the gluon--squark interaction
and the quartic term for the interaction between two gluons and two squarks] 
one also has a Yukawa--like term for the interaction of a quark, a squark and 
a gluino; SUSY imposes that the two coupling constants are the same $g_Y=g$. 

There is also a term for the self--interactions between the scalar fields; 
in the case where squarks have the same helicity and flavor, one has
\begin{eqnarray}
{\cal L}_{\rm self} = -g^2/3 \, ( \delta^{il} \delta^{kj} +
\delta^{ij} \delta^{kl}) \,  \phi_i \phi_j^* \phi_k \phi_l^*
\end{eqnarray}
Finally, there are the Yukawa interactions which generate the fermion 
masses, and the soft--SUSY breaking parameters which give masses to the gaugino
and scalar fields and introduce the trilinear couplings $A_q$. In the MSSM,
these terms can be written in a simplified way for the first generation as 
[$u$ and $d$ are the 
left--handed quarks, $\tilde{u}$ and $\tilde{d}$ their partners and $Q/
\tilde{Q}$ the left--handed doublets]  
\begin{eqnarray}
{\cal L}_{\rm Yuk} &=& h_u Q H_1 u^c +h_d Q H_2 d^c \\
{\cal L}_{\rm soft} &=& - m_{\tilde{g}}/2 \,  \bar{\lambda} \lambda
+ \sum m_{\tilde{q}_i}^2 \phi_i^* \phi_i  + \cdots 
+ h_u A_u \tilde{Q} H_1 \tilde{u}^c +h_d A_d \tilde{Q} H_2 \tilde{d}^c
+\cdots
\end{eqnarray}

\subsection{New features and complications compared to Standard QCD}

When one deals with calculations of QCD corrections in SUSY theories, 
a few complications compared to standard QCD corrections appear: 

-- Contrary to their standard partners the gluons, gluinos are massive 
particles due to the soft breaking of SUSY as discussed previously. 
In fact, gluinos are rather heavy in most of realistic and theoretically 
interesting models, and from the negative search of these states at the 
Tevatron a lower bound $m_{\tilde{g}} \gsim 200$ GeV has been set on their
masses; see Table 1. Light gluinos, which could be produced in 4--jet 
events at LEP1 seem to be experimentally ruled out \cite{pape}. Note 
also that gluinos are Majorana particles, and some care is needed in 
handling these states. 

-- As discussed previously, the left-- and right--handed current eigenstates 
$\tilde{q}_L$ and $\tilde{q}_R$, mix to give the mass eigenstates 
$\tilde{q}_1$ and $\tilde{q}_2$. The amount of mixing is 
proportional to the partner quark mass, and therefore is important only in 
the case of third generation, especially for the top squarks which can have
large mass splittings.  The mixing can also be important in the $\tilde{b}$ 
sector for large ${\rm tg}\beta$ values. 

-- In standard QCD, the only parameters are the QCD coupling constant 
$\alpha_s$ as well as the quark masses $m_q$ which in the high--energy 
limit can be set to zero. In SUSY--QCD, much more parameters are present: 
besides the $\tilde{q}$ masses [which are different in general]  and the 
$\tilde{g}$ mass, one has the soft--SUSY breaking trilinear couplings $A_q$ 
as well as the mixing angles $\theta_{\tilde{q}}$. These parameters are in 
general 
related, complicating the renormalisation procedure and making 
next--to--leading order calculations more involved since one has to deal 
with loop diagrams involving different particles or with multi-particle 
final states with several different masses.  

-- There is also a problem with the regularisation scheme. Indeed, the usual 
dimensional regularisation \cite{DREG} scheme which is used in standard QCD, 
breaks Supersymmetry \cite{DRED}. 
For instance the equality between the strong gauge coupling $g$ and the 
Yukawa coupling $g_Y$ is not automatically maintained at higher orders, and one
has to enforce it by adding additional counterterms. In the dimensional 
reduction scheme \cite{DRED0}, where only the four--vectors and not the Dirac 
algebra are 
in $n$--dimension, the equality between the two couplings is maintained 
automatically and this scheme is therefore more convenient. 
However, in some cases, gauge invariance can be broken in this scheme and 
again one has to add extra counterterms to satisfy the Ward identities.

-- Finally, there is an additional complication when Higgs bosons are involved. 
Indeed, when calculating QCD corrections for the pseudoscalar Higgs 
boson $A$, one has to be careful with the treatment of $\gamma_5$ beyond the 
one--loop level \cite{gamma5}. 

\subsection{Two--loop QCD corrections to the $\rho$ parameter} 

At ${\cal O}(\alpha \alpha_s)$, the two--loop Feynman diagrams contributing
to the $\rho$ parameter in SUSY consist of two sets which, at 
vanishing external momentum and after the inclusion of the counterterms, are 
separately ultraviolet finite and gauge-invariant. The first one has 
diagrams involving only gluon exchange, Fig.~4a; in this case 
the calculation is similar to the SM, although technically more complicated 
due to the larger number of diagrams and the presence of $\tilde{q}$ mixing. 
The diagrams involving the quartic scalar--quark interaction in Fig.~4a
will either contribute only to the longitudinal component of the 
self--energies or can be absorbed into the $\tilde{q}$ mass and mixing angle 
renormalisation as will be discussed later. The second set consists of 
diagrams involving scalar quarks, gluinos as well as quarks, 
Fig.~4b; in this case the calculation becomes very complicated due to 
the even larger number of diagrams and to the presence of up to 5 particles 
with different masses in the loops. 

\begin{figure}[htb]
\begin{center}
\mbox{
\psfig{figure=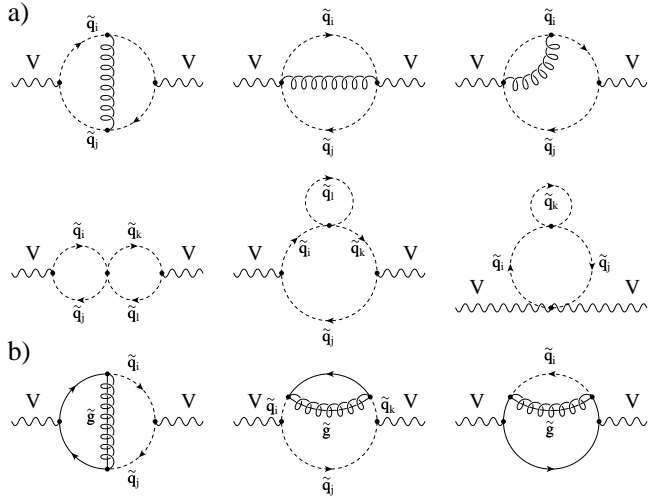,width=8cm,bbllx=165pt,bblly=495pt,bburx=450pt,bbury=725pt}}
\end{center}
\caption[]{Typical Feynman diagrams for the contribution of scalar 
quarks and gluinos to the $W/Z$--boson self--energies at the two--loop level.}
\vspace*{-4mm}
\end{figure}

The two--loop contribution of a complete quark/squark generation to the vacuum
polarization functions of the electroweak gauge bosons at zero
momentum--transfer have been calculated \cite{drho,drall} taking into account 
general mixing
between scalar quarks and allowing for all particles to have different masses.
The results were derived by two independent calculations using different
methods: one by evaluating the unrenormalised self-energies and the mass and
self-energy counterterms almost by hand and the other by using the package {\it
FeynArts} \cite{FA} where the relevant part of the MSSM has been 
implemented.  The two
independent calculations allowed for thorough checks of the final results.
  
The two--loop Feynman diagrams of Fig.~4 have to be supplemented by
the corresponding counterterm insertions into the one--loop diagrams. 
By virtue of the Ward identity, the vertex and wave--function
renormalisation constants cancel each other. The mass renormalisation
has been performed in the on--shell scheme, where the mass is
defined as the pole of the propagator. The mixing angle renormalisation 
is performed in such a way that all transitions from $\tilde{q}_i 
\leftrightarrow \tilde{q}_j$ which do not depend on the loop--momenta 
in the two--loop diagrams are canceled; this renormalisation condition 
is equivalent to the one used in Ref.~\cite{mixing} for scalar 
quark decays. With this choice of the mass and mixing angle renormalisation, 
the pure scalar quark diagrams in Fig.~4a that contribute to the 
transverse parts of the gauge--boson self--energies are canceled. 

In order to discuss the results, one can first concentrate on the 
contribution of the gluonic corrections, Fig.~4a, and the corresponding 
counterterms, which has a very simple analytical expression.  Indeed, 
the contribution of the $(\tilde{t}, 
\tilde{b})$ doublet to the $\rho$ parameter, including the two--loop 
gluon exchange and pure scalar quark diagrams is very simple if the
mixing in the $\tilde{b}$ sector is neglected; it is given by:
\begin{eqnarray}
\Delta \rho ^{\rm SUSY}_1 =\frac{G_F \alpha_s}{4 \sqrt{2} \pi^3} \left[ 
- s_t^2 c_t^2  
F_1\left( m_{\tilde{t}_1}^2,  m_{\tilde{t}_2}^2 \right) 
+ c_t^2 F_1 \left( m_{\tilde{t}_1}^2,  m_{\tilde{b}_L}^2 \right)
+s_t^2  F_1 \left( m_{\tilde{t}_2}^2,  m_{\tilde{b}_L}^2 \right) \right] 
\end{eqnarray}
where the two--loop function $F_1(x,y)$ is given in terms of dilogarithms by
\begin{eqnarray}
F_{1}(x,y) = a_+- 2\frac{xy}{a_-} \log \frac{x}{y} \left[2+
\frac{x}{y} \log \frac{x}{y} \right] 
+\frac{a_+x^2}{a_-^2}\log^2 \frac{x}{y} 
-2a_- {\rm Li}_2 \left(1-\frac{x}{y} \right) 
\end{eqnarray}
with $a_\pm=x\pm y$.  
This function is symmetric in the interchange of $x$ and $y$.
As in the case of the one--loop function $F_0$, it vanishes for 
degenerate masses, $F_1(x,x)=0$, while in the case of large 
mass splitting it increases with the heavy scalar quark mass 
squared: $F_1 (x,0) = x( 1 +\pi^2/3)$.  

The two--loop gluonic SUSY contribution to $\Delta \rho$ is shown in Fig.~5 
as a function of the common scalar mass $m_{\tilde{q}}$, for the two 
scenarios discussed previously: $\theta_{\tilde{t}} = 0$ and 
$\theta_{\tilde{t}} \simeq -\pi/4$. As can be seen, the two--loop contribution 
is of the order of 10 to 15\% of the one--loop result. Contrary to the SM 
case [and to many QCD corrections to electroweak processes in the SM, see
Ref.~\cite{BK} for a review] where the two--loop correction screens 
the one--loop contribution, $\Delta \rho_1^{\rm SUSY}$ has the same sign 
as $\Delta \rho_0^{\rm SUSY}$. For instance, in the case of degenerate 
$\tilde{t}$ quarks with masses $m_{\tilde{t}} \gg m_{\tilde{b}}$, the 
result is the same as the QCD correction to the $(t,b)$ contribution 
in the SM, but with opposite sign. The gluonic correction to the 
contribution of scalar quarks to the $\rho$ parameter will therefore 
enhance the sensitivity in the search of the virtual effects of scalar
quarks in high--precision electroweak measurements. 

\begin{figure}[htb]
\begin{center}
\mbox{
\psfig{figure=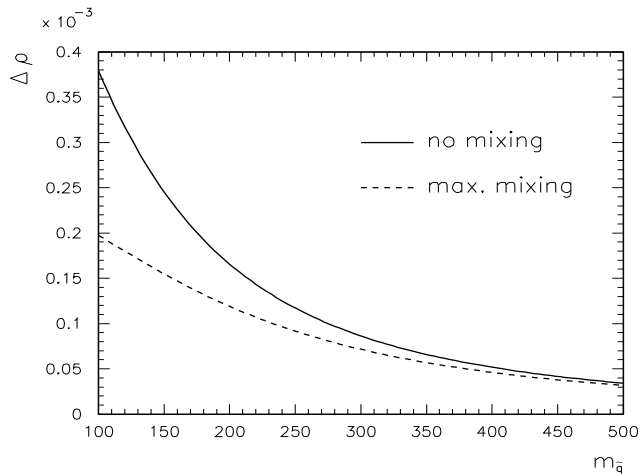,width=8cm,height=5.8cm,bbllx=140pt,bblly=285pt,bburx=450pt,bbury=535pt}}
\end{center}
\caption[]{Gluon exchange contribution to the $\rho$ parameter at two--loop 
as a function of $m_{\tilde{q}}$.}% for the scenarios of Fig.~2.}
\end{figure} 

The analytical expressions of the contribution of the two--loop diagrams 
with gluino exchange, Fig.~4b, to the electroweak gauge boson self--energies
are very complicated even at zero momentum--transfer. Besides the
fact that the scalar quark mixing leads to a large number of 
contributing diagrams, this is mainly due to the presence of up 
to five particles with different masses in the loops. The lengthy expressions 
are given in Ref.~\cite{drall}. It turned out that in general 
the gluino exchange diagrams give smaller contributions compared to gluon 
exchange.  Only for gluino and squark masses close to the 
experimental bounds they compete with the gluon exchange 
contributions. In this case, the gluon and gluino contributions add
up to $\sim 30\%$ of the one--loop value for maximal mixing; Fig.~6.
For larger values of $m_{\tilde{g}}$, the contribution
decreases rapidly since the gluinos decouple for high masses.

Finally, let us note that for the diagrams in Fig.~4a analytical
expressions for arbitrary momentum--transfer can be obtained as  
discussed in Ref.~\cite{drall}. With the present computational knowledge 
of two--loop radiative corrections, analytical exact results for the 
diagrams involving gluino exchange, Fig.~4b, cannot be obtained for 
arbitrary $q^2$; either approximations like heavy mass 
expansions or numerical methods have to be applied.

\begin{figure}[htb]
\begin{center}
\mbox{
\psfig{figure=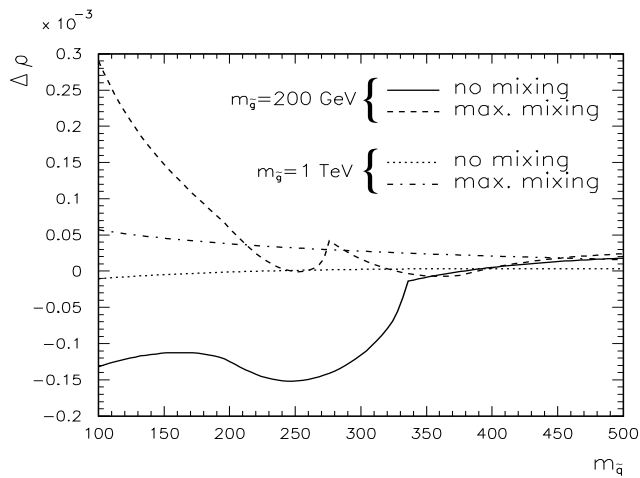,width=8cm,height=5.8cm,bbllx=140pt,bblly=285pt,bburx=450pt,bbury=535pt}}
\end{center}
\caption[]{Contribution of the gluino exchange diagrams to $\Delta \rho_1^
{\rm SUSY}$ for two values of $m_{\tilde{g}}$.}%  in the scenarios of Fig.~2.}
\end{figure} 

\vspace*{-3mm}

\section{Summary}

I have discussed the possibility of detecting the virtual effects of the new
particles predicted by supersymmetric extensions of the Standard Model in the
high--precision electroweak observables measured at LEP/SLC, the Tevatron and
CLEO. In view of the experimental accuracies on these observables and the
experimental limits on the SUSY particle masses, the two only observables where
these new effects might show up are the $\rho$ parameter and the radiative
decay $b \to s\gamma$. The experimental values of the two quantities agree
well, for the time being, with Standard Model expectations and rather strong
constraints on the MSSM parameter space can be obtained. In the not too far
future, more experimental accuracy can be achieved giving the hope that some
deviations from SM predictions might appear, thus showing for the first time an
indirect manifestation of SUSY.  

In order to make an accurate comparison between experimental and theoretical
values, high order effects must be included in the prediction for these two
observables. The next--to--leading order SUSY--QCD corrections have been made
available for both of them. In the last part of this lecture, I tried to
summarize the way, the techniques, and the complications of performing such
calculations, taking as an example the two--loop SUSY--QCD corrections to the
$\rho$ parameter.  

\section*{Acknowledgments:} 

I thank the organizers of the School, in particular Sergio Novaes and Rogerio 
Rosenfeld for their invitation and for the nice and lively atmosphere of the 
meeting. I also thank Manuel Drees for his hospitality and for being my 
companion in Sao Paulo.

\end{document}